\documentclass[aps, preprint, showpacs]{revtex4}
\begin{tiny}
\end{tiny}\begin{scriptsize}
\end{scriptsize}\topmargin -1cm
\usepackage{graphics}
\usepackage{graphicx}

\def \be {\begin{equation}}
\def \ee {\end{equation}}
\def \ba {\begin{eqnarray}}
\def \ea {\end{eqnarray}}
\def \bm {\begin{displaymath}}
\def \em {\end{displaymath}}
\def \br {{\bf r}}

\begin{document}
\title{ Freezing of a two dimensional fluid into a crystalline 
phase : Density functional approach }
\author {Anubha Jaiswal, Swarn L. Singh and Yashwant Singh}
\author{}
\affiliation{Department of Physics, Banaras Hindu University, 
Varanasi-221 005,
India}
\date{\today}
\begin{abstract}
A free-energy functional for a crystal proposed by Singh and Singh 
(Europhys. Lett. {\bf {88}}, 16005 (2009)) and which contains both the symmetry conserved
and symmetry broken parts of the direct pair correlation function has been used
to investigate the crystallization of a two-dimensional fluid. The results found 
for fluids interacting via the inverse power potential $ u(r)= \epsilon \left({\sigma}/{r} \right)^{n} $
for n= 3, 6 and 12 are in good agreement with experimental and 
simulation results. The contribution made by the 
symmetry broken part to the grand thermodynamic potential at 
the freezing point is found to increase with the softness of 
the potential. Our results explain why the Ramakrishnan-Yussouff 
(Phys. Rev. B {\bf 19}, 2775 (1979)) free-energy 
functional gave good account of freezing transitions of hard-core potentials 
but failed for potentials that have soft core and/or attractive tail.
\end{abstract}
\maketitle

\section{Introduction}
Freezing is a basic phenomenon, the most inevitable of all phase changes. 
When a liquid freezes into a crystalline solid its continuous symmetry  
of translation and rotation is broken into one of the Bravais lattices. A 
crystalline solid has a discrete set of vectors $\bf{ R}_{i} $ such that any
function of position such as one particle density $ \rho(\bf{r}) $ 
satisfies $ \rho (\bf{r} + \bf{R}_{i})=\rho(\bf{r}) $ 
for all $ \bf{R}_{i} $ [1]. Because of localization of particles on
lattice sites, a crystal is a system of extreme inhomogeneities where values of
$ \rho(\bf{r}) $ show extreme differences between its values on the 
lattice sites and in the interstitial regions. The density functional 
formalism of classical statistical mechanics has been used to develop
theories for liquid- solid transitions [2, 3]. This kind of approach was initiated 
in 1979 by Ramakrishnan and Yussouff [4].

         The central quantity in this formulation is the excess reduced
 Helmholtz free energy due to interparticle interactions of both the
 crystal $ A_{ex}[\rho] $ and the liquid $ A_{ex} [\rho_{l}] $ [4,5].
 For the crystal $ A_{ex}[\rho] $ is a unique functional of $ \rho{(\br)} $
 whereas for the liquid  $A_{ex}(\rho_{l}) $ is simply a function of liquid density
$ \rho_{l}$. The density functional formalism is used to write an expression 
for $ A_{ex}[\rho] $ in terms of one and two-particle distribution functions 
of the solid [2-5]. The direct pair correlation function (DPCF) that appears
in this expression is a functional of $\rho (\br)$ [2]. When this functional dependence 
is ignored and the  DPCF is replaced by that of the coexisting uniform liquid
[4,5] or by that of an `` effective uniform fluid " [6,7] the free energy functional
becomes approximate and fails to provide an accurate description of freezing 
transitions for a large class of intermolecular potentials. Attempts to include a 
term involving the three - body  direct correlation function of the coexisting liquid 
in the free energy functional failed to improve the situation [8, 9].

    It has recently been emphasized [10-12] that at the freezing point a
qualitatively new contribution to the correlations in distribution   
of particles emerges due to spontaneous symmetry breaking.
This fact has been used to write the DPCF of a frozen phase as
a sum of two qualitatively different contributions; one that preserves the continuous 
symmetry of uniform liquid and the other that breaks it and
 vanishes in the liquid. The double functional integration in density space of a 
 relation that connects $A_{ex} [\rho]$ 
to the DPCF led to an exact expression for $A_{ex} [\rho]$. 
The freezing transitions in three-dimensions have been investigated 
using this new free energy functional. The results found for the 
isotropic-nematic transition [10], crystallization of power-law fluids [11] and the
 freezing of fluids of hard spheres into crystalline and 
 glassy phases [12] are very encouraging.

      In this paper we apply the free energy functional to investigate the 
liquid-solid transition in two-dimensions. It may, however, be noted that in contrast 
to three-dimensional solid, a two-dimensional solid melts in two-steps; the 
intermediate phase known as hexatic has a very narrow stability region in between
liquid and crystal [13-15]. Since inclusion of the hexatic phase 
in the density functional formalism has not so for been possible, we 
neglect its presence and focus on the freezing of a fluid into the crystalline 
phase. Similar approach has been taken by others [16-21].        
Here our motivation is to examine how well this new free energy functional
 (described briefly in the following section)
compares with other free energy functionals in describing the crystallization  
of two-dimensional fluids.

     The paper is organized as follow : In Sec. II we give a brief description
of the free-energy functional for a symmetry broken phase that contains both the
symmetry-conserving and symmetry-broken parts of the DPCF. In Sec. III we describe 
methods to calculate these correlation functions for a two-dimensional system. The
theory is applied in Sec. IV to investigate the freezing of power-law fluids into
a crystalline solid of hexagonal lattice. The paper ends with a brief 
summary and perspectives in Sec. V.

 \section{Free Energy Functional}        
The reduced free energy functional $A[\rho]$ of an inhomogeneous system
is functional of $ \rho(\br)$ and is written as [2]

\begin{eqnarray}
A[\rho]= A_{id}[\rho] +  A_{ex} [\rho].    \nonumber \hspace*{5.5cm} (2.1)
\end{eqnarray}

The ideal gas part $ A_{id}$ is exactly known and is written in terms of
$ \rho(\br) $ as
\begin{eqnarray}
A_{id} [\rho] = \int{ d \br \rho (\br) [ ln(\rho (\br)\Lambda)- 1 ]},
 \nonumber \hspace*{4.cm} (2.2)
 \end{eqnarray}

where $\Lambda $ is a cube of the thermal wavelength associated with a molecule.
The second functional derivative of the excess part $ A_{ex} [\rho] $
with respect to $ \rho(\br) $  defines the DPCF
$ c (\br_{1}, \br_{2}) $ of the system [2],
\begin{eqnarray}
\frac{\delta^{2} A_{ex} [\rho]}{ \delta \rho(\br_{1})\delta\rho(\br_{2})} = 
-c (\br_{1},\br_{2}, [\rho] ). \nonumber \hspace*{5.cm} (2.3)
\end{eqnarray}

The function $ c $ that appears in this equation is related to the total correlation function
 $ h (\br_{1}, \br_{2}) $ through the Ornstein -Zernike(OZ) equation,
\begin{eqnarray}
h(\br_{1}, \br_{2}) = c(\br_{1},\br_{2}) + \int d \br_{3} c(\br_{1},\br_{3}) \rho (\br_{3})
h(\br_{2},\br_{3}) \nonumber . \hspace*{2.cm} (2.4)
\end{eqnarray}
Both functions h and c are functional of $ \rho(\br) $

    Since breaking of continuous symmetry  of a uniform liquid
at the freezing point gives rise to a qualitatively new contribution to
correlations in the distribution of particles [10-12], the DPCF of 
the frozen phase is written as a sum of two contributions;
\begin{eqnarray}
c(\br_{1},\br_{2};[\rho]) = c^{(0)}(|\br_{2}-\br_{1}|,\rho_{0}) + c^{(b)}(\br_{1},\br_{2};  [\rho]), \nonumber \hspace*{3cm} (2.5)
\end{eqnarray}
where $ c^{(0)}$ is symmetry-conserving and $c^{(b)}$ symmetry-broken parts of
the DPCF. While $ c^{(0)}$ depends on magnitude of interparticle separation r
and is function of average density $ \rho_{0}=\left\langle \rho (\br)\right\rangle $, 
$c^{(b)}$ is invariant only under discrete set of translations and rotations 
and is functional of $ \rho (\br) $. \\
 \hspace{1.cm}   Using Eq. $(2.5)$ we rewrite Eq. $(2.3)$ as
\begin{eqnarray}
\frac{\delta^{2} {A^{(0)}}_{ex}[\rho]}{\delta \rho(\br_{1})
\delta\rho(\br_{2} )} = - c^{(0)} (|\br_{2}- \br_{1}|,\rho_{0}), \nonumber \hspace*{5.5cm} (2.6)
\end{eqnarray}

\begin{eqnarray}
\frac{\delta^{2} {A^{(b)}}_{ex}[\rho]}{\delta \rho(\br_{1})
 \delta\rho(\br_{2} )}  = - c^{(b)} (\br_{1},\br_{2},[\rho]), \nonumber \hspace*{6.cm} (2.7)
\end{eqnarray}
where $ {A^{(0)}}_{ex} [\rho] +{A^{(b)}}_{ex} [\rho] = {A}_{ex} [\rho] $, The expressions
for $ {A^{(0)}}_{ex} [\rho] $ and $ {A^{(b)}}_{ex} [\rho] $ are found from functional
integrations of Eqs. $ (2.6) $ and $ (2.7)$, respectively. In this integration, as described
elsewhere [10-12], the system is taken from some initial density to 
the final density along a path in the density space; the
 result is independent of the path of integration.
 These integrations give,
\begin{eqnarray}
 {A^{(0)}}_{ex} [\rho] = {A}_{ex} (\rho_{l}) + \beta \mu - \ln {(\rho_{l}  \Lambda)}
 -{\frac{1}{2}} \int {d\br_{1} \int d \br_{2} (\rho(\br_{1})
-\rho_{l}) (\rho(\br_{2}) - \rho_{l}) \overline{c}^{(0)}(|\br_{2}-\br_{1}|)}
\nonumber , \hspace*{.5cm} (2.8) 
\end{eqnarray}
 and 
\begin{eqnarray}
 {A^{(b)}}_{ex} [\rho] = -\frac{1}{2} \int d\br_{1} \int d\br_{2} 
 (\rho(\br_{1}) - \rho_{0}) (\rho(\br_{2}) - \rho_{0}) \overline{c}^{(b)} (\br_{1},\br_{2})  \nonumber , \hspace*{4.cm} (2.9)
\end{eqnarray}
 where
\begin{eqnarray}
 \overline{c}^{(0)} (|\br_{2} -\br_{1}|) = 2 {{\int}_{0}}^{1} d\lambda \lambda
 {{\int}_{0}}^{1} d{\lambda}^{'} c^{(0)} (|\br_{2}-\br_{1}|, \rho_{l} + \lambda \lambda^{'}
 (\rho_{0}-\rho_{l})), \nonumber \hspace*{3.5cm} (2.10)
\end{eqnarray}
\begin{eqnarray}
 \overline{c}^{(b)} (\br_{1},\br_{2})= 4 {{\int}_{0}}^{1} d\xi \xi {{\int}_{0}}^{1} d\xi^{'}
 {{\int}_{0}}^{1} d\lambda \lambda {{\int}_{0}}^{1} d{\lambda}^{'} 
 c^{(b)} (\br_{1},\br_{2};\lambda \lambda^{'} \rho_{0}; \xi \xi^{'} \rho_{G}). \nonumber \hspace*{3cm} (2.11)
 \end{eqnarray}

Here $ A_{ex} (\rho_{l} )$ is excess reduced free energy
 of the coexisting uniform liquid of density $  \rho_{l} $ and 
chemical potential $ \mu $ , $ \rho_{0}= \rho_{l}(1 + \Delta \rho^{*}) $ 
is average density of the solid and $ \beta= k_{B} T$, $ k_{B}$ being 
the Boltzmann constant and T is the temperature and $ \rho_{G}$ is an 
order parameter arising due to breaking of symmetry.  

     The expression for $ {A_{ex}}^{(0)} [\rho]$ given by Eq. $ (2.8) $
is found from functional integrations when density $ \rho_{l} $
 of the coexisting fluid is taken as a reference.
The expression for $ {A_{ex}}^{(b)} [\rho]$ given by  Eq. $(2.9) $ is
found by performing double functional integrations in the density
space corresponding to the symmetry broken phase.
The path of integration in this space is characterised by two
 parameters $ \lambda $ and $ \xi $. These parameters vary
from $ 0 $ to $ 1 $. The parameter $ \lambda $
raises the density from zero to final value $ \rho_{_{0}} $
 as it varies from $ 0 $ to $ 1 $, whereas the parameter
 $ \xi $ raises the order parameters from zero to their final values $ \rho_{G} $.
The result is independent of the order of integration.
\\
\hspace{1.cm}   The free energy functional for the symmetry broken
phase is sum of $ A_{id}[\rho]$, ${A_{ex}}^{(0)}[\rho] $  and  
$ {A_{ex}}^{(b)}[\rho] $  Thus
 \begin{eqnarray}
 A [\rho]= \int d\br \rho(\br)  [ln (\rho (\br)\Lambda )- 1] + {A_{ex}}(\rho_{l}) +
\beta (\mu - ln(\rho_{l} \Lambda)) \int d\br (\rho (\br) -\rho_{l}) \nonumber \\ 
-\frac{1}{2} \int d \br_{1} \int d \br_{2} (\rho (\br_{1}) -\rho_{l})
(\rho (\br_{2}) -\rho_{l})
\overline{c}^{(0)} (|\br_{2} -\br_{1}|) \nonumber \hspace*{3.cm}  \\  
- \frac{1}{2} \int d \br_{1} \int d \br_{2}
  (\rho (\br_{1}) -\rho_{0}) (\rho (\br_{2}) -\rho_{0}) \overline{c}^{(b)}
 (|\br_{1},\br_{2}|) \nonumber \hspace*{2cm} (2.12)
\end{eqnarray} 
 where $ \overline{c}^{(0)} $ and $\overline{c}^{(b)} $ are given, respectively by
 $(2.10)$ and $(2.11) $.
 In deriving  Eq. $(2.12)$ no approximation has been introduced.
In the Ramakrishnan-Yussouff free energy functional  $ \overline{c}^{(b)} $ 
is neglected and $ \overline{c}^{(0)} $ is replaced by $ c^{(0)} $.\\
 In locating transition the grand thermodynamic potential defied as
 \begin{eqnarray}
 - W = A -\beta \mu \int d \br \rho(\br), \nonumber \hspace*{8.cm} (2.13)
  \end{eqnarray}
 is  generally used as it ensures that the pressure and the chemical potential
of the two phases remain equal at the transition. The transition point is
determined by the condition $ \Delta W= W_{l} - W =0$, where  $W_{l} $ is the grand
 thermodynamic potential of the coexisting liquid. From above expressions one
gets the following expression for $ \Delta W $;
\begin{eqnarray}
\Delta W =\int d \br [\rho (\br) \ln \frac{\rho (\br)}{\rho_{l}}
- (\rho (\br)- \rho_{l})]
-\frac{1}{2} \int d \br_{1} d \br_{2} (\rho (\br_{1})-\rho_{l})
(\rho (\br_{2})-\rho_{l})
{\bf \overline{c}^{(0)}} (|\br_{2}-\br_{1}|) \nonumber \\
-\frac{1}{2} \int d \br_{1} d \br_{2} 
(\rho (\br_{1})-\rho_{0})(\rho (\br_{2})-\rho_{0}) {\bf {\overline{c}}^{(b)}}
 (|\br_{1},\br_{2}|). \nonumber \hspace*{4.cm} (2.14)
 \end{eqnarray}
Minimization of $ \Delta W $ with respect to $ \rho (\br)$ subjected
 to the perfect crystal constraint leads to
\begin{eqnarray}
\ln \frac{\rho (\br)}{\rho_{l}} = \phi + \int d\br_{1} 
(\rho(\br_{2})-\rho_{l})\tilde{c}^{(0)}
(|\br_{2}- \br_{1}|) + \int d\br_{2} (\rho(\br_{1})-\rho_{0}) \tilde{c}^{(b)} 
(\br_{1},\br_{2}), \nonumber \hspace*{1.cm} (2.15)
\end{eqnarray}
  where
\begin{eqnarray}
\tilde{c}^{(0)} (|\br_{2}-\br_{1}|)= {\int_{0}}^{1} d\lambda {c}^{(0)} 
(|\br_{2}-\br_{1}|,\rho_{l}+\lambda (\rho_{0}-\rho_{l}) ), \nonumber 
\end{eqnarray}
and 
\begin{eqnarray}
\tilde{c}^{(b)} (\br_{1},\br_{2})= {\int_{0}}^{1} d\xi {\int_{0}}^{1}
 d\lambda {c}^{(b)} (\br_{1},\br_{2}, \lambda \rho_{0}, \xi \rho_{G} ). \nonumber 
 \end{eqnarray}
 The value of Lagrange multiplier $ \phi $ in Eq. $(2.15) $ is
 found from the condition
 \begin{eqnarray}
 \frac{1}{V} \int d \br \frac{\rho(\br)}{\rho_{0}} = 1 \nonumber \hspace*{6.cm} (2.16)
 \end{eqnarray}

      One needs, in principle, the values of $ c^{(0)} $ and $ c^{(b)} $ to
calculate self consistently  the value of $ \rho (\br) $ that
minimizes W. In practice however, one finds it convenient
 to do minimization with an assumed form of $ \rho(\br) $.
 The ideal part is calculated using a form for $ \rho(\br) $ which
is a superposition of normalized Gaussians centred around the lattice sites;
 \begin{eqnarray}
 \rho (\br) =\frac{\alpha}{\pi} \sum_{n} \exp [-\alpha ({\br- \bf{R}_{n}})^{2} ], \nonumber \hspace*{3.5cm} (2.17)
  \end{eqnarray}
 where $ \alpha $ is the localization parameter. For the interaction part it is
 convenient to use the Fourier expression,
\begin{eqnarray}
 \rho (\br) =\rho_{0} + \sum_{G \neq 0} \rho_{G} e^{i \bf{G}.\br}  \nonumber \hspace*{5.cm} (2.18)
\end{eqnarray}
where $ \bf {G} $ are reciprocal lattice vectors (RLV's) of the lattice
and $ \rho_{G}= \rho_{0} \mu_{G} $ are order parameters, Taking Fourier
transform of Eq. $ (2.17) $ one finds $ \mu_{G}= e^{(-G^{2}/4 \alpha)} $.

\section{Application to crystallization of power-law fluids}
 
 \subsection{Potential model}

We consider model fluids interacting via inverse power pair potentials
$ u(r)=\epsilon {(\sigma / r })^{n} $ ; where
 $ \epsilon, \sigma $
and n are potential parameters and r is molecular separation.
 The parameter n measures softness of the repulsion; $ n=\infty $
 corresponds to the hard disk and $ n=1 $ to the one component plasma. Such 
 repulsive potentials can be realised in colloidal suspensions. One such systems 
 in two-dimensions has been provided by paramagnetic
 colloidal particles in a pendant water droplet, which are confined
to the air- water interface [13]. By applying an external magnetic
field perpendicular to the interface, a magnetic moment is
introduced in the particles resulting into a tunable mutual
 dipolar repulsion between them. The pair interaction thus
created is repulsive and  proportional to $ r^{-3}  $. The
crystallization of this system has been investigated by
 van Teeffelen et.al. [20,21] using several versions of density
functional theory (DFT). Other example where short range repulsion
 between molecules is found is microgel spheres whose diameter
 could be temperature tuned [15]. Most computer simulation studies 
 on these systems suffer from the finite-size effects. In case of hard 
 disks recently a large scale 
Monte- Carlo simulation, large enough to access the thermodynamic
regime, has been performed [22]. The result confirms two-steps
transitions from liquid to solid with the intermediate haxatic 
phase[23,24]. However, the liquid- hexatic transition, in contrast 
to the prediction of Kosterlitz-Thouless-Halperin-Nelson-Young(KTHNY)
 theory [23,24] is found to be first-order while the hexatic-solid 
transition is second-order. The density functional theory predicts 
the liquid-solid transition to be first-order.

            In addition to being a pair potential that can be 
realized in a real system, it has a well known scaling property 
 according to which the reduced excess thermodynamic  properties depend
on a single variable (or coupling constant) which for a
two-dimensional  system is defined as
\begin{eqnarray}
\gamma = {(\rho \sigma^{2}) (\beta \epsilon)^{2/n} =\rho^{*} {T^{*}}^{-2/n}} \nonumber . 
\end{eqnarray}
  Using this scaling the potential is written as
\begin{eqnarray}
\beta u(r)= {\frac{\Gamma}{{r}^{n}} }  \nonumber   ,
\end{eqnarray}
where $ \Gamma = \gamma^{n/2}$ and r is measured in units of 
$ a_{0}={({1}/{\rho})}^{{1}/{2}} $.

\subsection{Calculation of $c^{(0)}(\br)$ and its derivatives with respect to $ \rho$ } 
The pair correlation functions of a classical system can 
be found in any spatial dimensions as a simultaneous solution 
of the OZ equation (Eq.$(2.4) $) and a closure relation that relates
functions h, c and the potential u(r). Several closure relations
including the Percus- Yevick (PY) relation, the hypernetted chain 
(HNC) relation, modified hypernetted chain (MHNC) relation 
etc. have been used to describe structure of a uniform fluid 
[25]. We may however note that while the OZ equation is general 
and connects the total and direct pair correlation functions
of liquids as well as of symmetry broken phases, the
closure relations that exist in the literature have been 
derived assuming translational invariance 
[25]. They are therefore valid only for normal fluids. We use the 
integral equation theory involving suitable closure relations 
to find symmetry-conserving part of pair correlation 
functions $ h^{(0)}(r)$ and $ c^{(0)} (r) $ and their derivatives 
with respect to density, The symmetry-broken part of the DPCF  
 is calculated using a method described in ref.[11].
  \\
\hspace*{1.cm}  The OZ equation for a uniform system of density $ \rho $ 
reduces to 
\begin{eqnarray}
h^{(0)}(r) = c^{(0)} (r) + \rho \int d \br^{'} c^{(0)}(r^{'}) 
h^{(0)} (|\br^{'}-\br|) \nonumber . \hspace*{5.5cm}  (3.1)
\end{eqnarray}
 The HNC closure relation and a closure relation proposed by Roger and
 Young [26] by mixing the PY and the HNC relations in such a way
 that at $ r= 0 $ it reduces to the PY and for $ r \rightarrow \infty $
it reduces to the HNC relation, can be written together as
\begin{eqnarray}
h^{(0)}(r)=  \exp (-\beta u(r)) \left[ 1+ \frac{ \exp \left\lbrace \chi (r)f(r)\right\rbrace 
- 1} {f(r)}\ \right]  -1 \nonumber  , \hspace*{3.cm}  (3.2)
\end{eqnarray}
 where $ \chi(r) =h^{(0)}(r)-c^{(0)}(r)  $ and $ f(r)= 1- e^{-\psi r}  $
 is a mixing function with an adjustable parameter $ 0 \leq \psi \leq \infty $.
 For $ \psi= \infty $ or, f(r)= $ 1 $, Eq. $ (3.2) $ reduces to the HNC closure 
 relation. In the Roger- Young relation, $ \psi$ is chosen to guarantee 
 thermodynamic consistency between the virial and compressibility 
 routes to the equation of state. 
\\ 
 \hspace*{1.cm}  The differentiation of Eqs. $ (3.1)$ and $ (3.2) $ with
  respect to density $ \rho $ yields following two relations 
\begin{eqnarray}
\frac{\partial h^{(0)} (r)}{\partial \rho} =
\frac{\partial c^{(0)} (r)}{\partial \rho} + \int d \br^{'} c^{(0)} (\br^{'})
h^{(0)} (|\br^{'}-\br|) +\rho \int d \br^{'} \frac {\partial c^{(0)} 
(\br)} { \partial \rho} h^{(0)} (|\br^{'}-\br|)\nonumber \\  
 +\rho \int d \br^{'} c^{(0)} (\br)
 \frac {\partial h^{(0)} (|\br^{'}-\br|)}  {\partial \rho} \nonumber \hspace*{6.cm}  (3.3)
 \end{eqnarray}  
and 
\begin{eqnarray}
\frac{\partial h^{(0)} (r)}{\partial \rho} = \exp {(-\beta u(r))}   
\exp [ \chi (r) f(r)] \frac{\partial \chi (r)}{ \partial \rho }
 {f^{-1}(r)}  \nonumber \hspace*{4.cm}  (3.4)
\end{eqnarray}
    The closed set of coupled equations $ (3.1) - (3.4) $ have been 
solved for four unknowns $ h^{(0)} $, $ c^{(0)} $ ,
$ \frac{\partial h^{(0)} (r)}{\partial \rho} $ and  
$ \frac{\partial c^{(0)} (r)}{\partial \rho} $.
The method can be extended to include higher order 
derivatives.
 In Fig. 1 we plot the Fourier transform of $ c^{(0)} (\br)$ defied as
\begin{eqnarray}
\hat{c}^{(0)} ({\bf{q}})= \rho \int d \br  c^{(0)} (\br) e^{i \bf{q}.\br} \nonumber , \hspace*{7.cm}  (3.5)
\end{eqnarray}

for $ \left( {n} ,  {\gamma} \right)$= $ \left( 3, 4.30\right)$, 
$ \left(6, 1.30 \right)$ and $ \left(12, 0.90 \right)$. 
The values given in Fig. $ 3.1 $ (a) for n=3 are in good agreement 
with values found by van Teeffelen et.al [20,21] (see Fig.1 of their 
paper). As has been reported in ref. [21] the HNC closure underestimates 
values of $ \hat{c}^{(0)} (q) $ whereas the Roger-Young (RY) closure 
gives relatively better but not very accurate values. In Fig. 1 we also 
give values found from an approach proposed by Kang and Ree (KR) [27].
\\  
 \hspace{0.5cm } The exact closure relation which one finds from the 
liquid state theory [25] can be written as 
\begin{eqnarray}
 1 + h^{(0)} (r)= g^{(0)} (r)= \exp  [-\beta u(r) + \chi (r) + B(r) ] \nonumber , \hspace*{3.cm}  (3.6)
 \end{eqnarray}
    where $ B(r)$ is the bridge function. In the HNC closure relation 
$ B(r)$ is taken equal to zero. In the KR approach the bridge function  
calculated for a reference potential and denoted as $ B_{0} (r)$ is used for
$ B(r)$ in Eq. $ (3.6) $. The evaluation of $ B_{0} (r)$  
is done prior to and separated from the main integral equation 
by solving the Martynov-Sarkisov [28] integral  equation. 
We briefly summarise here the way this is done for soft repulsive 
potentials in two-dimensions.
\\
         The potential $ u(r) $ is first divided into a reference 
$ u_{0}(r) $ and a perturbation part $ u_{p} (r) $.
\begin{eqnarray}
\hspace*{3.cm}u(r) = u_{0}(r) + u_{p} (r) \nonumber , \hspace*{5.cm}  (3.7)
\end{eqnarray}
where 
\begin{eqnarray}
u_{0} (r)= u (r) - F(r)   \hspace{1.cm}  if   \hspace{1.cm} r \leq a \nonumber \hspace*{4.cm} \\
\hspace*{2.4cm} =0 \hspace{2.6cm} if \hspace*{1.3cm} r >a  \nonumber \hspace*{3.2cm}  (3.8)
\end{eqnarray}
\begin{eqnarray}
u_{p} (r) = F(r) \hspace*{1.5cm} if \hspace{1.cm}  r \leq a \nonumber \hspace*{4.5cm} \\
\hspace*{2.5cm} =u (r) \hspace{1.5cm} if \hspace*{1.cm} r >a  \nonumber \hspace*{3.8cm}  (3.9)
\end{eqnarray}
     Here $ F(r)= u(a)- u^{'}(a) (a- r) $ and $ a $ is nearest neighbour
distance for hexagonal lattice at given density $ \rho $.
The $ B_{0} (r) $ for the reference potential is evaluated using 
the OZ equation 
\begin{eqnarray}
{h_{0}}^{(0)} (r)- {c_{0}}^{(0)} (r)= {\chi_{0}}^{(0)} (r) =\rho \int d \br^{'} {c_{0}}^{(0)} (r^{'}) {h_{0}}^{(0)} (|\br- \br^{'}|) \nonumber \hspace*{2.cm}  (3.10)
\end{eqnarray}
and closure relation 
\begin{eqnarray}
1+ {h_{0}}^{(0)} (r) = \exp [-\beta u_{0} (r)+\chi_{0} (r) +B_{0}(r) ] \nonumber . \hspace*{3.cm}  (3.11)
\end{eqnarray}
For $ B_{0}(r) $ the Mortynov- Sarkisov [28] relation 
\begin{eqnarray}
B_{0}(r) = [1+ s\chi_{0}(r) ]^{{1}/{s}} -1 -\chi_{0} (r) \nonumber , \hspace*{4.cm}  (3.12)
\end{eqnarray}
with $ s $=2 is used. The values of $ B_{0}(r) $ are found by solving 
Eqs.(3.10)-(3.12) self-consistently. The value of $ B_{0} (r) $
as a function of r for n= 3 is plotted in Fig. 2. The nature of $ B_{0} (r) $ 
is same as was found in case of  three-dimensions [27]. This value of $ B_{0} (r) $ found for 
the reference potentials is used in relation (3.6) which is 
used to solve the OZ equation self-consistently 
 to get values of ${h^{(0)}} (r) $ and $ {c^{(0)}} (r) $.
The values of $ {\widehat{c}^{(0)}}(q) $ found by this method are shown  
in Fig. 1 by full lines. These values are close to the simulation 
values given by van Teeffelen et.al [20,21] for n=3. For n=6 and 12 values found 
from the RY closure and values found from the KR method are close showing 
that for short-range repulsive potentials the RY closure yields 
good values of pair correlation functions.

\subsection{ Calculation of $ c^{(b)} (\br_{1}, \br_{2})$}

For a crystal $c^{(b)} (\br_{1}, \br_{2})  $ is invariant only under 
discrete set of translations corresponding to lattice vectors 
$ \lbrace \bf{R_{n}}\rbrace $. If one chooses a center of mass
variable $ \br_{c} =\frac{(\br_{1}+\br_{2})}{2} $ and difference variable 
$ \br =\br_{2}- \br_{1}  $, the $ c^{(b)} $ can be written as [11,12]
\begin{eqnarray}
c^{(b)} (\br_{1}, \br_{2})= \sum_{G} \exp (i {\bf{G}}.\br_{c}) c^{(G)}
 (\br;[\rho]) \nonumber , \hspace*{4.cm}  (3.13)
 \end{eqnarray}
where G are RLV's. Since $ c^{(b)} $ is real and symmetric with respect to 
interchange of $ \br_{1} $ and $ \br_{2} $ , $ c^{(G)}(\br)= c^{(-G)}(\br) $
and $ c^{(G)}(\br)= c^{(G)}(-\br) $. The function $ c^{(b)} (\br_{1}, \br_{2}) $ can
be expanded in terms of higher body direct correlation functions of 
uniform liquid [2];
\begin{eqnarray}
c^{(b)} (\br_{1}, \br_{2};[\rho])= \int d \br_{3} {c_{3}}^{(0)} (\br_{1},\br_{2},\br_{3};\rho_{0}) (\rho(\br_{3})- \rho_{0}) \nonumber  \hspace*{6.5cm} \\ 
 +{\frac{1}{2}} \int  d \br_{3} d \br_{4} {c_{4}}^{(0)} (\br_{1},\br_{2},
 \br_{3}, \br_{4};\rho_{0}) (\rho(\br_{3})- \rho_{0}) (\rho(\br_{4})- \rho_{0}) + \cdots \nonumber , \hspace*{1.cm}  (3.14)
 \end{eqnarray} 

where $ \rho(\br_{n}) -\rho_{0} = \sum_{G} \rho_{G} e^{i\bf{G}. \br_{n}}$,
and $ {c_{n}}^{(0)} $ are the n-body direct correlation functions of 
a uniform liquid of density $ \rho_{0} $. These correlation functions are 
related to derivatives of $ c^{(0)}(r,\rho_{0}) $ with respect to density
$ \rho_{0} $ as follows [2] ;
\begin{eqnarray}
\frac{\partial c^{(0)}(r)}{\partial\rho_{0}}= \int d \br_{3} 
{c_{3}}^{(0)} (\br_{1},\br_{2},\br_{3}) \nonumber , \hspace*{3.5cm}
\end{eqnarray}
\begin{eqnarray}
\frac{\partial^{2} c^{(0)}(r)}{\partial{\rho_{0}}^{2}}= \int d \br_{3} 
\int d \br_{4}{c_{4}}^{(0)} (\br_{1},\br_{2},\br_{3},\br_{4}) \nonumber , \hspace*{1.cm}  (3.15)
\end{eqnarray}
etc. \\
      The values of derivatives of $ c^{(0)} (r) $ appearing on 
the left hand side of above equations  can be found using the integral 
equation theory described above. The usefulness of this method to find 
$ c^{(b)}(\br_{1},\br_{2}) $ depends on convergence of series Eq. $(3.14)$  
which is a series in ascending powers of order parameters, 
and our ability of finding values of n-body $ (n \geq 3)$ direct correlation 
functions from Eq. $ (3.15) $. Barrat et.al [8]
have shown that $ {c_{3}}^{(0)} $ can be factored as
$ {c_{3}}^{(0)} (\br_{1},\br_{2},\br_{3}) = t(r_{12}) 
t(r_{13})t(r_{23})$ and the function t(r) can be determined from 
relation (see Eq.  $ (3.15) $)
\begin{eqnarray}
\frac{\partial c^{(0)}(r)}{\partial\rho_{0}}= t(r) \int d \br^{'}  
t(r^{'}) t(|\br^{'}-\br|) \nonumber. \hspace*{2.cm}  (3.16)
\end{eqnarray}
This method can be extended for higher $ {c_{n}}^{(0)} $ [29].
Since $\overline{c}^{(b)} (\br_{1}, \br_{2}) $ is averaged over density
$ \rho $ and over order parameters $ \rho_{G} $ the contributions made
by successive terms of Eq. $(3.14) $ in $ {A_{ex}}^{(b)} [\rho] $
is expected to decrease rapidly [11]. In the case of three 
dimensions it was found that it is only first term of the series $ (3.14) $
which needs to be considered to describe accurately 
the fluid -solid transition [11, 29]. In two-dimensions the
convergence is expected to be faster as number of nearest neighbours
is less compared to the three-dimensions and therefore the higher 
body correlation functions expected to be less important. In view of this,
 we consider here the first term of the series $(3.14) $ only and examine its contribution
 in stabilizing the hexagonal lattice at the transition point.
 
       From known values of $ \frac{\partial c^{^{(0)}(r)}}{\partial \rho} $
we solved numerically Eq. $ (3.16) $ to find values of t(r)
for different values of $ \gamma $. In Fig. 3, we plot values
of t(r) for n = $ 3, 6 $ and $ 12 $ at values of $ \gamma $ 
close to the freezing point.
\\
   Taking only the first term of Eq. $ (3.14) $ the expression for
$ c^{(b)} (\br_{1}, \br_{2}) $ in terms of t(r) can be written as 
\begin{eqnarray}
c^{(b)} (\br_{1}, \br_{2})= \rho_{0} \sum_{G} \int d \br_{3}
t(|\br_{3}- \br_{1}|) {e^{i\bf {G}.\br_{3}}} t(|\br_{3}- \br_{2}|)\nonumber . \hspace*{2.cm}  (3.17)
\end{eqnarray}
Using the relation 
\begin{eqnarray}
\br_{3} = \frac{1}{2} {(\br_{1}+ \br_{2})} +(\br_{3}-\br_{1})
-\frac{1}{2} (\br_{2}- \br_{1}) \nonumber \\
= \br_{c} +\br^{'}- \frac{1}{2} (\br_{2}- \br ) \nonumber , \hspace*{2.6cm} 
\end{eqnarray}

Eq. $ (3.17) $ reduces to Eq. $(3.13) $, i.e.
\begin{eqnarray}
c^{(b)} (\br_{1},\br_{2})= \sum_{G} e^{i\bf{G}. \br_{c}} c^{(G)}(\br) \nonumber ,
\end{eqnarray}  
where 
\begin{eqnarray}
c^{(G)}(\br)= \rho_{0} \mu_{G} t(r) e^{-\frac{1}{2} i\bf {G}.\br}
\int d \br^{'} t(r^{'}) e^{ i\bf {G}.\br^{'}} t(|\br^{'}-\br|) \nonumber . \hspace*{2.cm} (3.18)
\end{eqnarray}
Using the relation $ e^{ i\bf {G}.\br}= \sum_{m} (i)^{m} J_{m}(Gr)
e^{im (\phi_{G}-\phi_{r})} $ where $ J_{m}(Gr) $ is the 
Bessel function of the first kind of integral order m
we find following expression for $ c^{(G)} (r) $;
\begin{eqnarray}
c^{(G)}(r)= \sum_{M} (i)^{M} {c_{M}}^{(G)}(r) e^{-iM \phi_{r}} \nonumber , \hspace*{5.cm}  (3.19)
\end{eqnarray}
where
\begin{eqnarray}
{c_{M}}^{(G)}(r)= \rho_{0} \mu_{G} t(r) \sum _{m} B_{m} (r,G)
J_{m+M}{ \left( \frac{1}{2} Gr \right) }e^{i M \phi_{G}}  \nonumber \hspace*{2.cm} (3.20)
\end{eqnarray}
and
\begin{eqnarray}
B_{m}(r,G)= \int  dk  k t(k)  J_{m}(kr) e^{im \phi_{r}} 
 \int dr^{'} r^{'} J_{m}(kr^{'}) J_{m} (G r^{'}) \nonumber \hspace*{1.5cm} (3.21)
\end{eqnarray}
For hexagonal lattice $ M=0,\pm 6  $. The value of $ {c_{M}}^{(G)}(r) $
depends on values of order parameters $ \mu_{G} $ and on the values of 
RLV's. In Figs. 4-6, we plot harmonic coefficients $ {c_{0}}^{(G)}(r) $ 
and $ {c_{6}}^{(G)} (r) $ for $ n=3, 6 $ and $ 12 $ for RLV's of first four sets,
 respectively. For different set of RLV's $ {c_{M}}^{(G)}(r) $ varies with r in
different way; the values in all cases become negligible for r
( measured in units of $ a_{0} = {(1/ \rho)}^{1/2} $)$  > 1.5$ .
For any given value of G, the values of $ {c_{0}}^{(G)}(r) $ is about an order 
of magnitude larger than $ {c_{6}}^{(G)}(r) $ at their maxima and minima. 
As magnitude of G increases value of $ {c_{M}}^{(G)} (r) $ 
decreases and after the sixth set of RLV's values of 
$ {c_{M}}^{(G)} (r) $ become negligible for all values of n.

\section{Liquid- Solid transition}

   Substituting expressions of $ \rho (r) $ given by Eqs. $ (2.17) $
and $ (2.18) $ and of $ \overline{c}^{(b)} (\br_{1}, \br_{2}) $
given by Eq. $ (3.19) $ in Eq. $ (2.14) $ we 
find.
\begin{eqnarray}
\frac{\Delta W}{N} = \frac{\Delta W_{id}}{N} +\frac{\Delta W_{0}}{N} + 
\frac{\Delta W_{b}}{N} \nonumber , \hspace*{3.cm}  (4.1)
\end{eqnarray}
where
\begin{eqnarray}
\frac{\Delta W_{id}}{N} = (1 + \Delta \gamma) [1 + \ln  \left( \frac{\alpha}{\pi} \right)  -2 -
\ln \rho_{0}] +1 \nonumber , \hspace*{1.cm}  (4.2)  \\ 
\frac{\Delta W_{0}}{N} = - \frac{1}{2} \Delta {\gamma} c^{(0)}(0) - \frac{1}{2}
\sum_{G \neq 0 } {|\mu_{G}|}^{2} \widehat{c}^{(0)} (G) 
 \nonumber, \hspace*{1.5cm}  (4.3)\\ 
\frac{\Delta W_{b}}{N} = - \frac{1}{2} \sum_{G} {\sum_{G_{1}}}^{'} \mu_{G_{1}} 
\mu_{-G-G_{1}} \hat{\overline{c}} ({\bf{G}_{1} + \frac{1}{2} \bf{G}}) \nonumber. \hspace*{1.5cm}  (4.4)
\end{eqnarray}
 where $ \Delta \gamma=  \left( \frac{\gamma_{s}- \gamma_{l}}{\gamma_{l}} \right)  $ ;
 the subscripts s and l represent solid and liquid respectively. 

Hear $\Delta W_{id}, \Delta W_{0}  $ and $ \Delta W_{b} $ are respectively,
the ideal, symmetry-conserving and symmetry-broken contributions to
$ \Delta W $. The prime on a summation in Eq $ (4.4) $ indicates the condition
$   {\bf{G}} \neq 0, {\bf{G}}_{1} \neq 0 $ and ${ \bf{G}}+
 {\bf{G}}_{1} \neq 0 $, and 
 \begin{eqnarray}
\hat{c}^{(0)} (G) = \int c^{(0)} (r) e^{i \bf{G}.\br} d\br \nonumber \hspace*{2.cm}  \\ 
 \hat{\overline{c}} ({\bf{G}_{1} + \frac{1}{2} \bf{G}}) =  
\int  \overline{c}^{(G)} (r;\rho_{0}) e^{-i ({\bf{G}_{1}+\frac{1}{2} \bf{G})}. 
{\br}}  d \br  \nonumber  
 \end{eqnarray}
 
      We used above expressions to locate the liquid-crystal (hexagonal lattice)
transition by varying values of $ \gamma, \Delta \gamma $  and $ \alpha $.
The results given in Table 1 for n= $ 3, 6 $ and $ 12 $ correspond to the RY closure 
relation. We note that the contribution arising due to symmetry broken part of the
DPCF is far from negligible and its importance increases with the softness of 
potential. While it is about  7.3 $\% $  to the symmetry conserving term for 
 n=12, it increases about  44 $\% $ for n=3. This explains why the 
Ramakrishnan-Yussouff theory gives good results for hard core potentials 
but fails for potentials that have soft core and/or attractive tail. As
the contribution of $ \frac{\Delta W_{b}}{N} $ is negative, it stabilizes 
the solid phase. Without it the theory strongly overestimates the stability 
of fluid phase specially  for softer potentials [20,21].
The contribution made by the symmetry broken part of the DPCF is, as expected, 
small compered to that in three-dimensions (3D) at the freezing point for the same 
potential. For example, the contribution in 3D [11] for n=12 is 22.2 $ \% $
compared to 7.3 $ \% $  in  2D  whereas for n=6 the contribution is 
 37 $ \% $ in 3D and  18 $ \% $ in 2D .

    In Table 2 we compare results of the present calculation using 
both the RY closure and the KR procedure to calculate pair correlation 
functions for n=3 with the results found from other free-energy functionals as 
reported in ref $[21]$. The experimental results obtained from real-space
microscopy measurements of magnetic colloids confined to an air-water interface
[13] and values found from numerical simulations [30,31]
are also given in the table. While the RY closure gives slightly higher 
values of $ \Gamma_{f} $ and $ \Gamma_{s} $ compared to experimental values,
the KR closure gives slightly lower values. But these values along with the 
values of other parameters, particularly the values of $ \Delta \Gamma= \Gamma_{s}- \Gamma_{f} $,
are in better agreement with the experimental values compared to any other 
versions of the DFT. Although the extended modified weighted-density approximation 
(EMA) [32] with Verlet closure [33] gives values of $ \Gamma_{f} $ which is close to 
the one found by us using the KR closure, but the values of $ \Delta \Gamma $ 
is significantly lower; $ \Delta {\Gamma}_{EMA}=0.16 $ compared to the values found by us
$ \Delta\Gamma=0.41 $, and the experimental value, $ 0.75 $ .

    The real-space experimental data are not available for other systems.
The computer simulation results [34, 35] show the liquid-solid transition 
at $ \gamma_{l}= 1.51 $ and $ 0.986 $ respectively for $ n=6 $ and $ 12 $. 
These values are close to the one given in Table 1.

\section{Summery and Perspectives }
    We used a free energy functional that contains both the symmetry-conserving 
part of the DPCF $ c^{(0)} (r) $ and the symmetry-broken part 
$ c^{(b)} (\br_{1},\br_{2}) $ to investigate the freezing of a two-dimensional
fluid into a two-dimensional crystal of hexagonal lattice. The values of $ c^{(0)} (r) $ and 
its derivatives with respect to density $ \rho $ as a function of interparticle 
separation r have been determined using an integral equation theory comprising   
the OZ equation and the closure relations of Roger and Young [26] and of Kang and Ree
[27]. For soft potential (n=3) the two results are found to differ; the KR closure seems 
to give better result. For more repulsive potentials the two results are close 
as shown in Fig. 1. For $ c^{(b)} (\br_{1},\br_{2}) $ which is functional of 
$ \rho (\br) $ and is invariant only under discrete set of translations and 
rotations, we used an expansion in ascending powers of order parameters. This 
expansion involves higher body direct correlation functions of isotropic  
phase, which in turn were found from the density derivatives of $ c^{(0)} (r) $ 
using a method proposed by Barrat et.al [8].

   The contribution of symmetry-broken part of DPCF to the free energy is found to depend on  
nature of pair potentials; the contribution increases with softness 
of potentials. This result is in agreement with that found in three-dimensions
and explains why the Ramakrishnan-Yussouff free-energy functional was 
found to give a reasonably good description of the freezing transition of 
hard core potentials but failed for potentials that have soft core and/or
attractive tail. The results found here and the results reported for 
3D indicate that the theory described here can be used to investigate the 
freezing transitions of all kinds of fluids.  

    Since our free energy functional takes into account the spontaneous 
symmetry breaking it can be used to study various phenomena of ordered 
phases. The results indicate that the density-functional approach 
provides an effective frame work for theoretical study of a large 
variety of problems involving inhomogeneities. However, the question not 
adequately addressed yet is the size of fluctuations effect which play 
important role in two-dimensional systems. The other 
important question is the inclusion of hexatic phase in the theory.

{\bf{Acknowledgments}}: We are thankful to J. Ram for 
computational help. One of us (Anubha) is thankful to the University 
Grants Commission for research fellowship.


\newpage 
\vspace{.5in}
Table 1. Freezing parameters $ \alpha, \gamma_{l}, \Delta \gamma  $ and 
the pressure P at coexistence along with the contributions of
ideal, symmetry-conserving and symmetry-broken parts of 
$ \frac{\Delta W}{N}  $. These results correspond to the
Roger-Young closure [26].

\begin{table}[h]
\begin{center}
\begin{tabular}{|c|c|c|c|c|c|c|c|} \hline
\hspace{0.4cm} n \hspace{0.4cm}  &\hspace{0.4cm} $ \alpha $ \hspace{0.4cm} &
\hspace{0.4cm} $ \gamma_{l} $ \hspace{0.4cm}   & \hspace{0.4cm} $ \Delta \gamma $ \hspace{0.4cm} 
 & \hspace{0.4cm} $ \frac{\Delta W_{id} }{N} $  \hspace{0.4cm} & 
\hspace{0.4cm} $ \frac{\Delta W_{0} }{N} $ \hspace{0.4cm} & \hspace{0.4cm} $  \frac{\Delta W_{b} }{N} $ \hspace{0.4cm}
&\hspace{0.4cm} $ \frac{\beta P}{\rho}  $  \hspace{0.4cm}  \\  \hline
 3  & 100 & 4.96  & 0.025 & 2.50 & -1.74 & - 0.76 & 75 \\  \hline 
 6  & 100 & 1.55  & 0.040 & 2.50 & -2.10 & - 0.40 & 31 \\  \hline 
 12 & 96 & 1.00  & 0.050 & 2.49 & -2.32 & - 0.17 & 22 \\  \hline 
\end{tabular}
\end{center}
\end{table}

Table 2. Freezing parameters $ \Gamma_{f} (= {\gamma_{f}}^{n/2}) $,
$ \Gamma_{s} (= {\gamma_{s}}^{n/2} )$ and the width of coexistence region
$ \Delta \Gamma= \Gamma_{s} -\Gamma_{f} $, and the relative displacement
parameter $ \xi (\simeq 2/ \alpha) $  at the coexistence obtained from 
various density functional schemes. The MWDA stands for modified 
weighted density approximation, EMA for extended modified weighted 
density approximation, RY and KR refer to, respectively, the Roger-Young
closure [26] and the Kang and Ree [27] closure.

\begin{table}[h]
\begin{center}
\begin{tabular}{|c|c|c|c|c|} \hline
  & \hspace{0.5cm} $ \Gamma_{l} $ \hspace{0.5cm}  & \hspace{0.5cm} $ \Gamma_{s}   $ 
 \hspace{0.5cm} &  \hspace{0.5cm} $\Delta \Gamma $ \hspace{0.5cm}   
   &  \hspace{0.5cm} $ \xi $   \hspace{0.5cm}  \\  \hline
Present result with RY &  11.04 &  11.46 & 0.42 & 0.020   \\ 
Present result with KR &  9.20  &    9.61 & 0.41 & 0.022 \\ \hline 

MWDA with RY  [21]   &    41.07 &   41.13 &   0.06 & 0.017  \\ 
 EMA with RY    [21]  &    23.00 &   23.08 &   0.09 & 0.020   \\ 
EMA with Verlet [21] &    9.33  &   9.49  &   0.16 & 0.020  \\ \hline

 Simulation [31]  & 12.0 & 12.25 & 0.025 & -    \\ 
 Experiment [13] & 10.0 & 10.75 & 0.75 & 0.038  \\ \hline

\end{tabular}
\end{center}
\end{table}
\newpage
\begin{figure}[ht]
\includegraphics[height=7.0in, width=4.0in,clip]{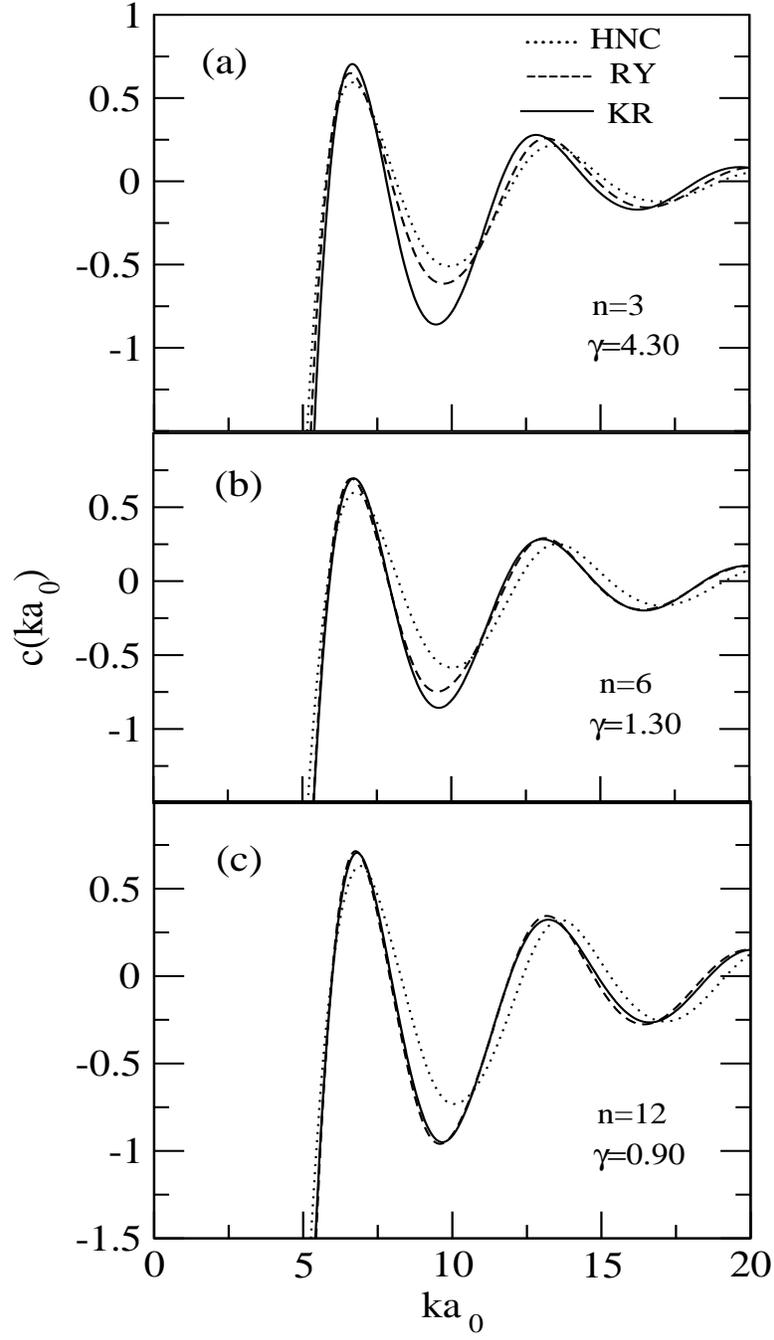}
\caption {The dimensionless Fourier transform $ \hat{c}^{(0)} (q) $ 
of the direct pair correlation function of $ {c}^{(0)} (r) $ plotted
against $ ka_{0}  \left( a_{0}= \left( 1/ \rho \right)^{1/2} \right) $ 
shown are data found from the integral equation theory using the RY 
closure (dashed line), HNC closure (dotted line) and KR closure 
(full line) at values of $ \gamma $ shown in figure (a), (b) 
and (c) for n=3, 6 and 12, respectively.}
\end{figure}

\begin{figure}[ht]
\includegraphics[height=3.0in, width=4.0in,clip]{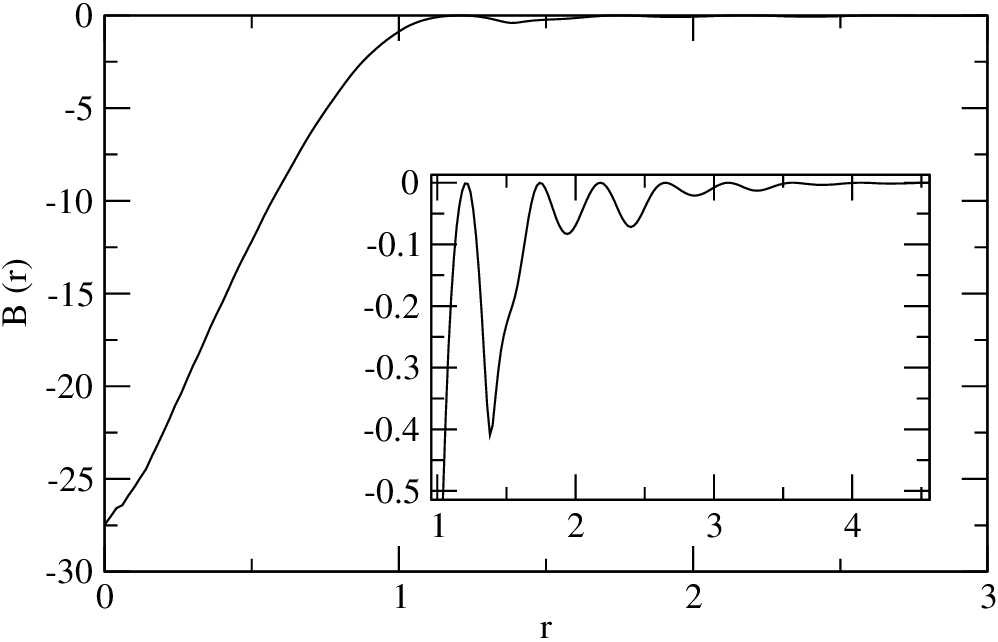}
\caption{ Bridge function $ B_{0} (r) $ for n=3 at $ \gamma= 4.30 $.
 The distance $ r $ is in units of $ a_{0}={(1/\rho)}^{1/2} $.
 Inset magnifies the values of $ B_{0} (r) $ for $ r \geq 1 $.}
\end{figure}

\begin{figure}[ht]
\includegraphics[height=3.0in, width=4.0in,clip]{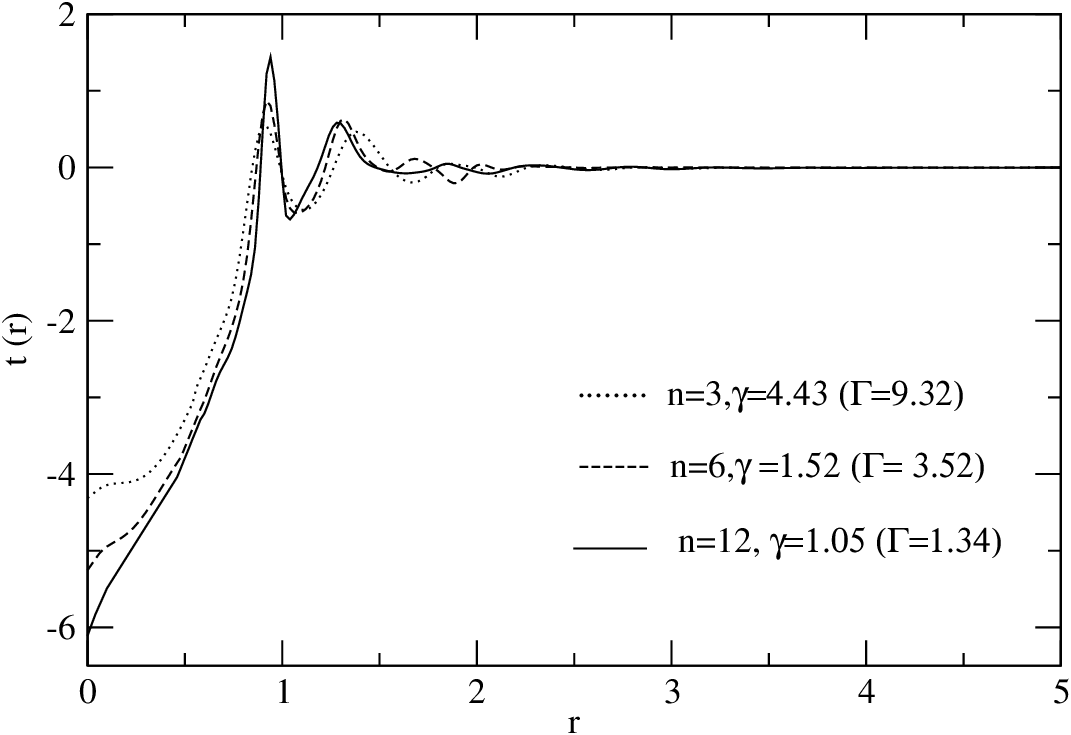}
\caption {Function t(r) vs r for n=3, 6 and 12  at the values 
of $ \gamma $ above the fluid-solid transition point, r is in 
 units of $ a_{0}= \left({1}/{\rho} \right)^{1/2} $.}  
\end{figure}

\begin{figure}[ht]
\includegraphics[height=3.6in, width=4.0in,clip]{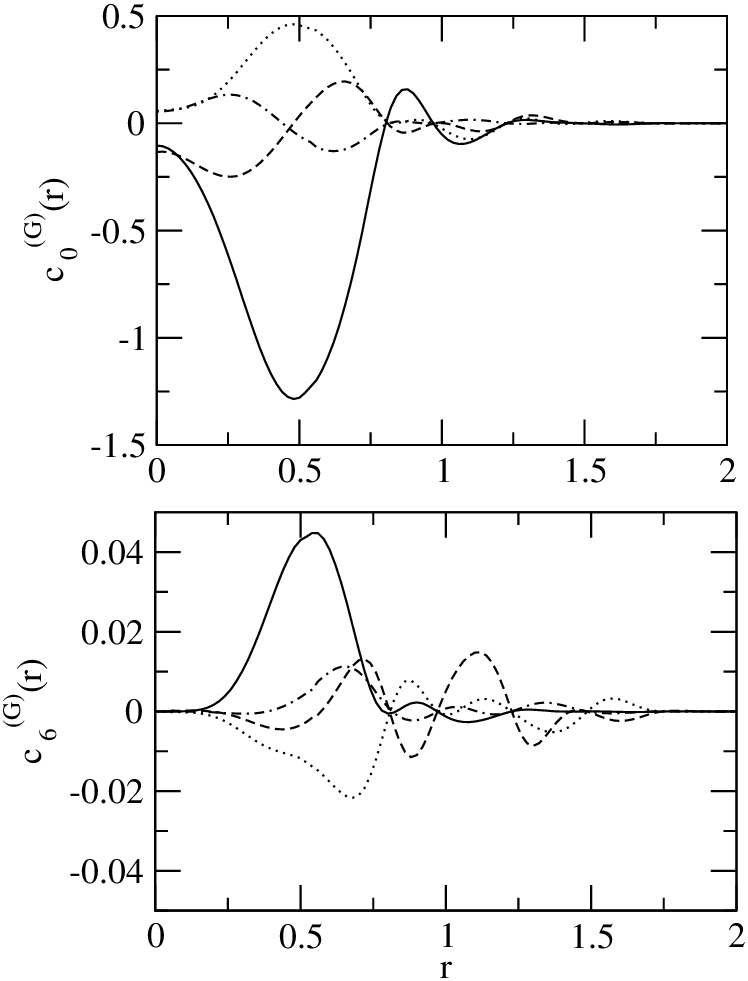}
\caption {Harmonic coefficients ${c_{M}}^{(G)} (r) $
for RLV's of first four sets for n=3, $\gamma=4.42 (\Gamma=9.32) $.
Notations are as follow: Full line represents values of the first
set, the dotted line of the second set , dashed line of the third 
set and dashed-dotted line of the fourth set. The distance r is 
expressed in unit $ a_{0} $, where $ a_{0} ( ={ {1}/{\rho}})^{{1}/{2}} $.}
\end{figure}

\begin{figure}[ht]
\includegraphics[height=3.6in, width=4.0in,clip]{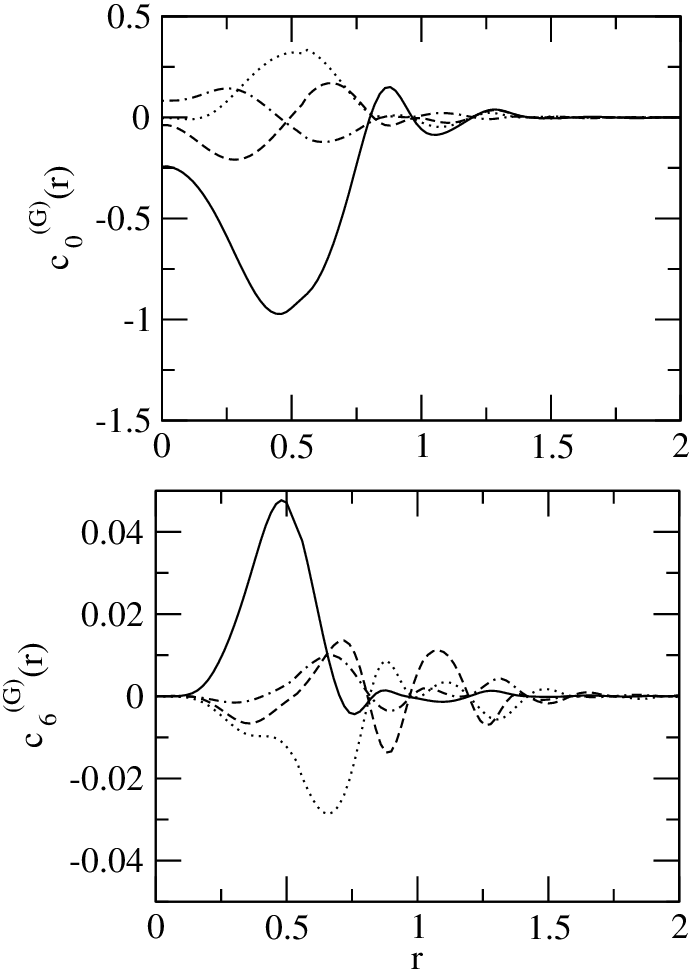}
\caption {Harmonic coefficients ${c_{M}}^{(G)} (r) $
for RLV's of first four sets for n=6, $\gamma=1.52 (\Gamma=3.52) $. Notations are same as in Fig 4.}.
\end{figure}

\begin{figure}[ht]
\includegraphics[height=3.6in, width=4.0in,clip]{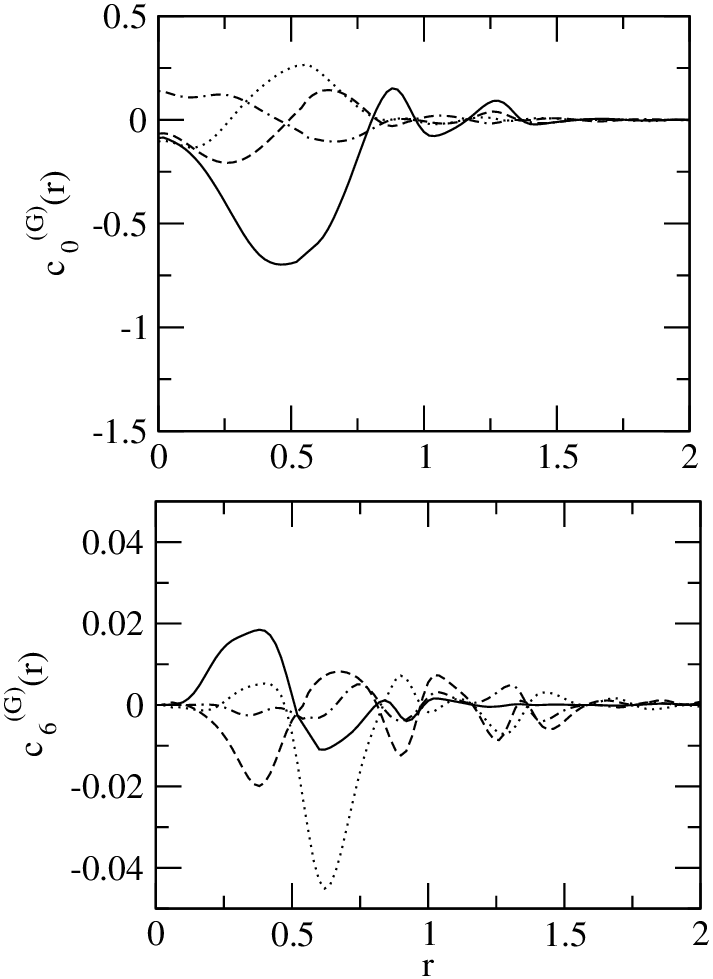}
\caption {Harmonic coefficients ${c_{M}}^{(G)} (r) $
for RLV's of first four sets for n=12, $\gamma=1.05 (\Gamma=1.34) $.
Notations are same as in Fig 4.}
\end{figure}


\end{document}